\documentclass[prd,nofootinbib,preprint]{revtex4}

\usepackage{amsmath,amssymb}
\usepackage[normalem]{ulem}
\usepackage{graphicx}
\usepackage{array}

\newcommand{\eVq}{\ensuremath{\text{eV}^2}}
\newcommand{\Dmq}{\Delta m^2}
\newcommand{\Dlt}{\Delta\delta}
\newcommand{\Eps}{{\varepsilon}}
\newcommand{\Epp}{{\varepsilon'}}

\newcolumntype{C}{>{~$}c<{$~}}
\newcolumntype{R}{>{~$}r<{$~}}

\begin{document}

\title{Atmospheric Neutrino Oscillations and New Physics}

\author{M.C.~Gonzalez-Garcia}
\email{concha@insti.physics.sunysb.edu}
\affiliation{%
  C.N.~Yang Institute for Theoretical Physics,
  SUNY at Stony Brook, Stony Brook, NY 11794-3840, USA
  \\
  IFIC, Universitat de Val\`encia - C.S.I.C., Apt 22085, 
  E-46071 Val\`encia, Spain}

\author{Michele Maltoni}
\email{maltoni@insti.physics.sunysb.edu}
\affiliation{%
  C.N.~Yang Institute for Theoretical Physics,
  SUNY at Stony Brook, Stony Brook, NY 11794-3840, USA}

\begin{abstract}
    \vspace*{1cm}%
    We study the robustness of the determination of the neutrino
    masses and mixing from the analysis of atmospheric and K2K data
    under the presence of different forms of phenomenologically
    allowed new physics in the $\nu_\mu$--$\nu_\tau$ sector. We focus
    on vector and tensor-like new physics interactions which allow us
    to treat, in a model independent way, effects due to the violation
    of the equivalence principle, violations of the Lorentz invariance
    both CPT conserving and CPT violating, non-universal couplings to
    a torsion field and non-standard neutrino interactions with
    matter.  We perform a global analysis of the full atmospheric data
    from SKI together with long baseline K2K data in the presence of
    $\nu_\mu \to \nu_\tau$ transitions driven by neutrino masses and
    mixing together with sub-dominant effects due to these forms of
    new physics. We show that within the present degree of
    experimental precision, the extracted values of masses and mixing
    are robust under those effects and we derive the upper bounds on
    the possible strength of these new interactions in the
    $\nu_\mu$--$\nu_\tau$ sector.
\end{abstract}

\preprint{YITP-SB-18-04}

\maketitle

\section{Introduction}

Neutrino oscillations are entering a new era in which the observations
from underground experiments obtained with neutrino beams provided to
us by Nature --~either from the Sun or from the interactions of cosmic
rays in the upper atmosphere~-- are being confirmed by experiments
using ``man-made'' neutrinos from accelerators and nuclear
reactors~\cite{review}.

For atmospheric neutrinos, Super--Kamiokande (SK) high statistics 
data~\cite{skatmlast,skatmpub} established beyond doubt that the observed 
deficit in the $\mu$-like
atmospheric events is due to the neutrinos arriving in the detector at
large zenith angles, and it is best explained by $\nu_\mu$ oscillations. 
This evidence was also confirmed by other atmospheric
experiments such as MACRO~\cite{macro} and Soudan 2~\cite{soudan}.

The KEK to Kamioka long-baseline neutrino oscillation experiment (K2K)
uses an accelerator-produced neutrino beam mostly consisting of
$\nu_\mu$ with a mean energy of 1.3~GeV and a neutrino flight distance
of 250~km to probe the same oscillations that were explored with
atmospheric neutrinos.  Their results~\cite{k2kprl} show that both the
number of observed neutrino events and the observed energy spectrum
are consistent with neutrino oscillations, with oscillation parameters
consistent with the ones suggested by atmospheric neutrinos.

Oscillations are not the only possible mechanism for atmospheric
$\nu_\mu \to \nu_\tau$ flavour transitions. They can also be generated
by a variety of forms of nonstandard neutrino interactions or
properties. In general these alternative mechanisms share a common
feature: they require the existence of an interaction (other than the
neutrino mass terms) that can mix neutrino flavours~\cite{npreview}.
Among others this effect can arise due to violations of the
equivalence principle (VEP)~\cite{VEP,VEP1,qVEP}, non-standard
neutrino interactions with matter~\cite{NSI}, neutrino couplings to
space-time torsion fields~\cite{torsion}, violations of Lorentz
invariance (VLI)~\cite{VLI1,VLI2} and of CPT
symmetry~\cite{VLICPT1,VLICPT2}. From the point of view of neutrino
oscillation phenomenology, the most relevant feature of these
scenarios is that, in general, they imply a departure from the
$E^{-1}$ energy dependence of the oscillation
wavelength~\cite{yasuda1}.

Prior to the highest-statistics SK data, some of these scenarios could
provide a good description --~alternative to $\Dmq$ neutrino
oscillations~-- of the atmospheric neutrino
phenomenology~\cite{oldatmfitnp,NSI2}. However, with more precise
data, and in particular with the expansion of the energy range covered
by atmospheric neutrino data due to the inclusion of the upward-going
muons, these alternative scenarios became disfavoured as leading
mechanism to explain the observations~\cite{fogli1,lipari,NSI3}.  The
results from K2K experiment further singled out oscillations as the
dominant mechanism of $\nu_\mu\leftrightarrow\nu_\tau$
transitions~\cite{fogli2}.\footnote{Recently~\cite{noon04} SK
  collaboration has presented a reanalysis of the SK1 data in terms of
  the reconstructed $L/E$ which allowed them to slightly improve the
  discrimination between oscillations and alternative mechanisms.
  Unfortunately, to reproduce such analysis for the subdominant effects
  discussed here is not possible outside the collaboration.}

The question arises, however, to what point the possible presence of
these forms of new physics, even if sub-dominant, may affect the
derived ranges of masses and mixing from the oscillation analysis of
the atmospheric and K2K data. Or in other words, to what level our
present determination of the neutrino masses and mixing is robust
under the presence of phenomenologically allowed new physics effects.

In this paper we address this question by performing a global analysis
of the atmospheric and K2K data with $\nu_\mu \to \nu_\tau$
transitions driven by neutrino masses and mixing in the presence of
some generic forms of new physics. In particular we consider new
physics interactions which are vector-like, or tensor-like (scalar
interactions cannot be distinguish from oscillations). This allow us to
treat, in a model independent way, effects due to the violation of the
equivalence principle, violations of the Lorentz invariance both CPT
conserving and CPT violating, non-universal couplings to a torsion
field and non-standard neutrino interactions in matter. 
In Sec.~\ref{sec:forma} we present the formalism adopted and
the data set used. In Sec.~\ref{sec:results} we show the results of
our analysis. Conclusions are given in Sec.~\ref{sec:conclu}. The
technical details of our new statistical analysis of the atmospheric
data are described in the appendix.

\section{Formalism}
\label{sec:forma}

In what follows we consider some new physics (NP) scenarios which
induce new sources of lepton flavour mixing in addition to the
``standard'' $\Dmq$ oscillations ($\Dmq$-OSC). We
concentrate on flavour mixing mechanisms for which the propagation of
neutrinos ($+$) and antineutrinos ($-$) is governed by the following
Hamiltonian~\cite{VLICPT2}:
\begin{equation} \label{eq:hamil}
    \mathbf{H}_\pm \equiv
    \dfrac{\Dmq}{4 E}
    \mathbf{U}_\theta
    \begin{pmatrix}
	-1 & ~0 \\
	\hphantom{-}0 & ~1
    \end{pmatrix}
    \mathbf{U}_\theta^\dagger
    + \sum_n
    \sigma_n^\pm \dfrac{\Dlt_n\, E^n}{2}
    \mathbf{U}_{\xi_n,\pm\eta_n}
    \begin{pmatrix}
	-1 & ~0 \\
	\hphantom{-}0 & ~1
    \end{pmatrix}
    \mathbf{U}_{\xi_n,\pm\eta_n}^\dagger \;,
\end{equation}
where $\Dmq$ is the mass--squared difference between the two neutrino
mass eigenstates, $\sigma_n^\pm$ accounts for a possible relative
sign of the NP effects between neutrinos and antineutrinos and
$\Dlt_n$ parametrizes the size of the NP terms. The matrices
$\mathbf{U}_\theta$ and $\mathbf{U}_{\xi_n,\pm\eta_n}$ are given by:
\begin{equation} \label{eq:rotat}
    \mathbf{U}_\theta =
    \begin{pmatrix}
	\hphantom{-}\cos\theta & ~\sin\theta \\
	-\sin\theta & ~\cos\theta
    \end{pmatrix}\,,
    \qquad
    \mathbf{U}_{\xi_n,\pm\eta_n} =
    \begin{pmatrix}
	\hphantom{-}\cos\xi_n\hphantom{e^{-i\eta_n}} 
	& ~\sin\xi_n e^{\pm i\eta_n} 
	\\
	-\sin\xi_n e^{\mp i\eta_n} 
	& ~\cos\xi_n\hphantom{e^{-i\eta_n}}
    \end{pmatrix}\,,
\end{equation}
where we have also accounted for possible non-vanishing relative
phases $\eta_n$. For concreteness we will focus on NP effects which
are induced by tensor-like and vector-like interactions.

We denote by tensor-like interactions those with $n=1$ leading to a
contribution to the oscillation wavelength which grows linearly with
the neutrino energy. For example, Eq.~\eqref{eq:hamil} can describe
the evolution of $\nu_\mu$ and $\nu_\tau$'s of different masses in the
presence of violation of the equivalence principle (VEP) due to non
universal coupling of the neutrinos, $\gamma_1\neq \gamma_2$ ($\nu_1$
and $\nu_2$ being related to $\nu_\mu$ and $\nu_\tau$ by a rotation
$\theta_G$), to the local gravitational potential
$\phi$~\cite{VEP,VEP1}.\footnote{VEP for massive neutrinos due to
  quantum effects discussed in Ref.~\cite{qVEP} can also be parametrized
  as Eq.~(\ref{eq:hamil}) with $n=2$.} Phenomenology of neutrino
oscillations induced or modified by VEP has been widely studied in the
literature~\cite{VEPpheno}.

In this case
\begin{equation} \label{eq:veq}
    \Dlt_1 = 2 |\phi|(\gamma_1- \gamma_2) \equiv 2 |\phi| \Delta\gamma \,,
    \qquad \xi_1 = \theta_G \,,
    \qquad \sigma_1^+ = \sigma_1^- \,.
\end{equation}
For constant potential $\phi$, this mechanism is phenomenologically
equivalent to the breakdown of Lorentz invariance induced by different
asymptotic values of the velocity of the neutrinos, $v_1\neq v_2$,
with $\nu_1$ and $\nu_2$ being related to $\nu_\mu$ and $\nu_\tau$ by
a rotation $\theta_v$~\cite{VLI1,VLI2}. In this case
\begin{equation} \label{eq:vli}
    \Dlt_1 = (v_1- v_2)\,\delta v \,,
    \qquad \xi_1 = \theta_v \,,
    \qquad \sigma_1^+ = \sigma_1^- \,.
\end{equation}

We denote by vector-like interactions those with $n=0$ which induce an
energy independent contribution to the oscillation wavelength. This
may arise, for instance, from a non-universal coupling of the
neutrinos, $k_1\neq k_2$ ($\nu_1$ and $\nu_2$ being related to the
$\nu_\mu$ and $\nu_\tau$ by a rotation $\theta_Q$), to a space-time
torsion field $Q$~\cite{torsion}, so
\begin{equation} \label{eq:torsion}
    \Dlt_0= Q (k_1- k_2)\equiv Q \,\delta k \,,
    \qquad \xi_0 = \theta_Q \,,
    \qquad \sigma_0^+ = \sigma_0^- \,.
\end{equation}
Violation of CPT due to Lorentz-violating effects also lead to an
energy independent contribution to the oscillation
wavelength~\cite{VLICPT1,VLICPT2} with
\begin{equation} \label{eq:cpt}
    \Dlt_0 = b_1-b_2 \equiv \delta b \,,
    \qquad \xi_0 = \theta_{\not\text{CPT}} \,,
    \qquad \sigma_0^+ = -\sigma_0^-
\end{equation}
where $b_i$ are the eigenvalues of the Lorentz violating CPT-odd
operator $\bar{\nu}_L^\alpha b_\mu^{\alpha\beta} \gamma_\mu
\nu_L^\beta$ and $\theta_v$ is the rotation angle between the
corresponding neutrino eigenstates and the flavour
eigenstates~\cite{VLICPT2}.

In all these scenarios, if the NP strength is constant along the
neutrino trajectory, the expression of $P_{\nu_\mu \to\nu_\mu}$
takes the form~\cite{VLICPT2}:
\begin{equation} \label{eq:prob}
    P_{\nu_\mu \to \nu_\mu} = 1 - P_{\nu_\mu \to \nu_\tau} =
    1 - \sin^2 2\Theta \, \sin^2 \left( 
    \frac{\Dmq L}{4E} \, \mathcal{R} \right) \,.
\end{equation}
where the correction to the $\Dmq$-OSC wavelength, $\mathcal{R}$, and
to the global mixing angle, $\Theta$, verify
\begin{align}
    \mathcal{R} \cos 2\Theta \,
    & = \cos 2\theta + \sum_n\, R_n\, \cos 2\xi_n \,,
    \\
    \mathcal{R} \sin 2\Theta \
    & = \lvert \sin 2\theta 
    + \sum_n\, R_n \,\sin 2\xi_n \, e^{i\eta_n} \rvert \,,
\end{align}
with $R_n$ being the ratio between $\Dmq$--induced and the NP--induced
contributions to the oscillation wavelength
\begin{equation}
    R_n = \sigma_n^+ \frac{\Dlt_n E^n}{2} \, \frac{4E}{\Dmq} \,.
\end{equation}
For $P_{\bar{\nu}_\mu \to \bar{\nu}_\mu}$ the same expressions hold
with the exchange $\sigma_n^+ \to \sigma_n^-$ and $\eta_n \to
-\eta_n$.

For the sake of simplicity, in what follows we concentrate in
scenarios with one NP source characterized by a unique $n$. In this
case
\begin{align}
    \label{eq:Theta}
    \sin^2 2\Theta &= \frac{1}{\mathcal{R}^2} \left(
    \sin^2 2\theta + R_n^2 \sin^2 2\xi_n
    + 2 R_n \sin 2\theta \sin 2\xi_n \cos\eta_n \right) \,,
    \\[2mm]
    \label{eq:Xi}
    \mathcal{R} &= \sqrt{1 + R_n^2 + 2 R_n \left( \cos 2\theta \cos 2\xi_n
      + \sin 2\theta \sin 2\xi_n \cos\eta_n \right)} \;.
\end{align}

\begin{figure}
    \includegraphics[width=4.5in]{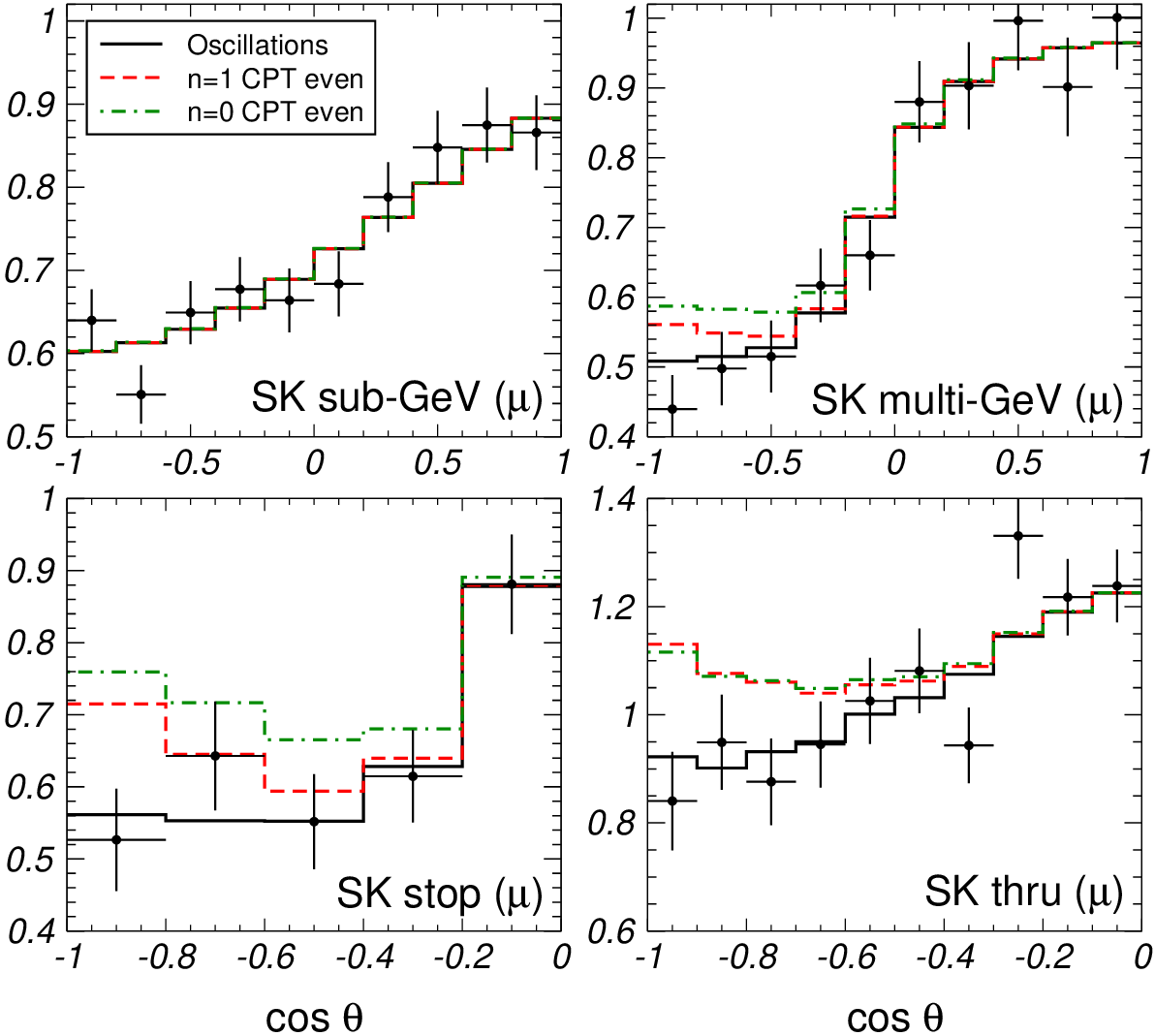}
    \caption{\label{fig:zenith}%
      Zenith-angle distributions (normalized to the no-oscillation
      prediction) for the Super--Kamiokande $\mu$-like events. The
      full line gives the distribution for the best fit of $\Dmq$-OSC,
      $\Dmq = 2.2\times 10^{-3}~\eVq$ and $\sin^2\theta = 0.5$. The
      dashed and dotted lines give the distributions for $\Dmq$-OSC+NP
      scenarios for $n=1$ and $n=0$ with $\Dlt_1 = 2.0\times 10^{-24}$
      and $\Dlt_0 = 4.2\times 10^{-23}$~GeV respectively. In both
      cases $\eta=\xi=0$ and the oscillation parameters have been set
      to their best fit values.}
\end{figure}

In Fig.~\ref{fig:zenith} we illustrate the effect of the presence of
the NP in the atmospheric neutrino events distributions for $\Dmq$-OSC
plus sub-dominant CPT-even tensor-like and vector-like NP effects, for
some characteristic values of the NP-parameters. In both cases $R_n$
is a growing function of $E$ and the NP effects become relevant in the
higher energy samples, in particular for upward going muons.

In order to properly define the intervals of variation of the five
parameters $\Dmq$, $\theta$, $\Dlt_n$, $\xi_n$, $\eta_n$, we can take
advantage of the symmetries of the Hamiltonian (see also
Ref.~\cite{Fornengo:2001pm} for a very similar problem). For a given
value of $\sigma_n^+$, from the expressions~\eqref{eq:hamil}
and~\eqref{eq:rotat} we see that the Hamiltonian is invariant under
the following transformations:
\begin{itemize}
  \item $\theta \to \theta + \pi$,
  \item $\xi_n \to \xi_n + \pi$,
  \item $\eta_n \to \eta_n + 2\pi$,
  \item $\Dmq \to -\Dmq \quad \text{and} \quad
    \theta \to \theta + \pi/2$,
  \item $\Dlt_n \to -\Dlt_n \quad \text{and} \quad
    \xi_n \to \xi_n + \pi/2$,
  \item $\xi_n \to -\xi_n \quad \text{and} \quad \eta_n \to \eta_n + \pi$.
\end{itemize}
Furthermore, the relevant survival probabilities $P_{\nu_\mu \to
  \nu_\mu}$ and $P_{\bar{\nu}_\mu \to \bar{\nu}_\mu}$ are not affected
by a change in the overall sign of the Hamiltonian, as well as change
in the global phase of its non-diagonal components. Therefore, we also
have:
\begin{itemize}
  \item $\theta \to \theta + \pi/2 \quad \text{and} \quad
    \xi_n \to \xi_n + \pi/2$,
  \item $\theta \to -\theta \quad \text{and} \quad
    \xi_n \to - \xi_n$,
  \item $\eta_n \to -\eta_n$.
\end{itemize}
The above set of symmetries allows us to define the ranges of
variation of the five parameters as follows:
\begin{align}
    (a) &~ \Dmq \geq 0 \,, &
    (c) &~ 0 \leq \theta \leq \pi/2 \,, \nonumber 
    \\
    (b) &~ \Dlt_n \geq 0 \,, &
    (d) &~ 0 \leq \xi_n \leq \pi/4 \,, \label{eq:intervals}
    \\
    && (e) &~ 0 \leq \eta_n \leq \pi \,. \nonumber
\end{align}
Thus in the general case we cover the mixing parameter space by using,
for instance, $0 \leq \sin^2\theta\leq 1$ and $0 \leq \sin^22\xi_n
\leq 1$.

For the case of real relative phase, $\eta_n \in \{0,\,\pi\}$, one can
absorb the two values of $\eta_n$ into the sign of $\xi_n$. In this
case we drop $(e)$ and replace $(d)$ by:
\begin{equation} \label{eq:xietazer0}
    (d')\; -\pi/4 \leq \xi_n \leq \pi/4
\end{equation}
and use instead  $-1\leq\sin 2\xi_n\leq 1$.

Finally we notice that the above derivation is valid for a given sign
of $\sigma_n^+$. Keeping the convention of $\Dmq > 0 $ and $\Dlt_n >
0$ the survival probability for the opposite sign can be obtained by
the exchange
\begin{equation} \label{eq:sign}
    \sin^2\theta \to 1-\sin^2\theta
    \quad \text{and} \quad
    \eta_n \to \pi - \eta_n \,.
\end{equation}

In addition, we also consider the special case of vector-like NP due
to non-standard neutrino-matter interactions (NSI)~\cite{NSI,NSI2}.
In this case the effective Lagrangian describing the evolution of the
$\nu_\mu$--$\nu_\tau$ system can be written
as~\cite{NSI2,Fornengo:2001pm}
\begin{equation} \label{eq:halnsi}
    \mathbf{H}_\pm =
    \dfrac{\Dmq}{4 E} \mathbf{U}_\theta
    \begin{pmatrix}
	-1 & ~0 \\
	\hphantom{-}0 & ~1
    \end{pmatrix}
    \mathbf{U}_\theta^\dagger
    \pm \sqrt{2} \, G_F N_f(r)
    \begin{pmatrix}
	-\Epp_\pm/2 & \Eps_\pm \hphantom{/2} \\
	\hphantom{-}\Eps^*_\pm \hphantom{/2} & \Epp_\pm/2
    \end{pmatrix}
\end{equation}
where $N_f(r)$ is the number density of the fermion $f$ along the path
$\vec r$ of the neutrinos propagating in the Earth, and $\Eps_\pm$ and
$\Epp_\pm$ parametrize the deviation from standard neutrino
interactions: $\sqrt{2} \, G_F N_f(r) \, \Eps_+$ is the forward
scattering amplitude of the FC process $\nu_\mu + f \to \nu_\tau + f$,
and $\sqrt{2} \, G_F N_f(r) \, \Epp_+$ is the difference between the
$\nu_\tau + f$ and the $\nu_\mu + f$ elastic forward scattering
amplitudes. The corresponding amplitudes for antineutrinos are given
by $\Eps_-$ and $\Epp_-$. For simplicity we assume that the
interactions for neutrinos and antineutrinos are the same, which
implies $\Epp_+ = \Epp_{-} \equiv \Epp$ and $\Eps_+ = \Eps^*_{-}
\equiv \Eps$. Thus the NSI Hamiltonian contains three real parameters,
which can be chosen to be $\Epp$, $|\Eps|$ and $\arg(\Eps)$. NSI and
their interplay with the oscillations have also been studied in
different contexts: among others, in relation to supernova
neutrinos~\cite{NSISN}, to the solar neutrino problem~\cite{NSIsolar},
to the LSND results oscillation results~\cite{NSILSND}, to
neutrinoless double beta decay~\cite{NSIbeta}, and to present and
future laboratory neutrinos~\cite{NSIlab}.

Formally Eq.~\eqref{eq:halnsi} can be seen as a special case of 
Eq.~\eqref{eq:hamil} with $n=0$, $\sigma_0^- = -\sigma_0^+$, and
\begin{gather} \label{eq:parnsi} 
    \Dlt_0 = 2\sqrt{2}\, G_F \, N_f(\vec{r})\, {\cal F}
    \equiv 4.58 \times 10^{-22}\, (2-Y_p)\,
    \frac{\rho(\vec{r})_\text{Earth}}{3 \rm g/cm^3} \, 
    {\cal F}\; \text{GeV} \,, \nonumber
    \\[2mm]
    \cos(2\xi) = \frac{\Epp/2}{\cal F} \,, \quad 
    \sin(2\xi) = \frac{|\Eps|}{\cal F} \,, \quad 
    \eta = \arg(\Eps) \,,
    \\[2mm]
    \text{with} \quad {\cal F} =\sqrt{|\Eps|^2 + \frac{\Epp^2}{4} } \;.
    \nonumber
\end{gather}
Technically the main difference is that NSI only affect the evolution
of neutrinos in the Earth, and their strength changes along the
neutrino trajectory. Consequently the flavour transition probability
cannot be simply read from Eq.~\eqref{eq:prob} and its evaluation
requires the numerical solution of the neutrino evolution in the Earth
matter.  In our calculations we use PREM~\cite{PREM} for the Earth's
density profile and a chemical composition with proton/nucleon ratio
$Y_p=0.497$ in the mantle and 0.468 in the core.  In what follows for
the sake of concreteness we set our normalization on these parameters
by considering that the relevant neutrino interaction in the Earth
occurs only with down--type quarks.

Concerning the data samples used in the analysis, for atmospheric
neutrinos we include in our analysis all the contained events as well
as the the upward-going neutrino-induced muon fluxes from the latest
1489 SK data set~\cite{skatmlast}. This amounts for a total of 55 data
points. Details of our new statistical analysis, introduced here for
the first time, are presented in the appendix.

For the analysis of K2K we include the data on the normalization and
shape of the spectrum of single-ring $\mu$-like events as a function
of the reconstructed neutrino energy~\cite{k2kprl}. We bin the data in
five 0.5~GeV bins with $0 < E_\text{rec} < 2.5$~GeV plus one bin
containing all events above 2.5~GeV. Details of our analysis of the
K2K data were presented in Ref.~\cite{ourthree} and will not be
repeated here. Let us just comment that together with statistical
uncertainties we also account for the systematic uncertainties
associated with the determination of the neutrino energy spectrum in
the near detector, the model dependence of the amount of nQE
contamination, the near/far extrapolation and the overall flux
normalization.

\section{Results and Discussion}
\label{sec:results}

\begin{figure}
    \includegraphics[width=4.5in]{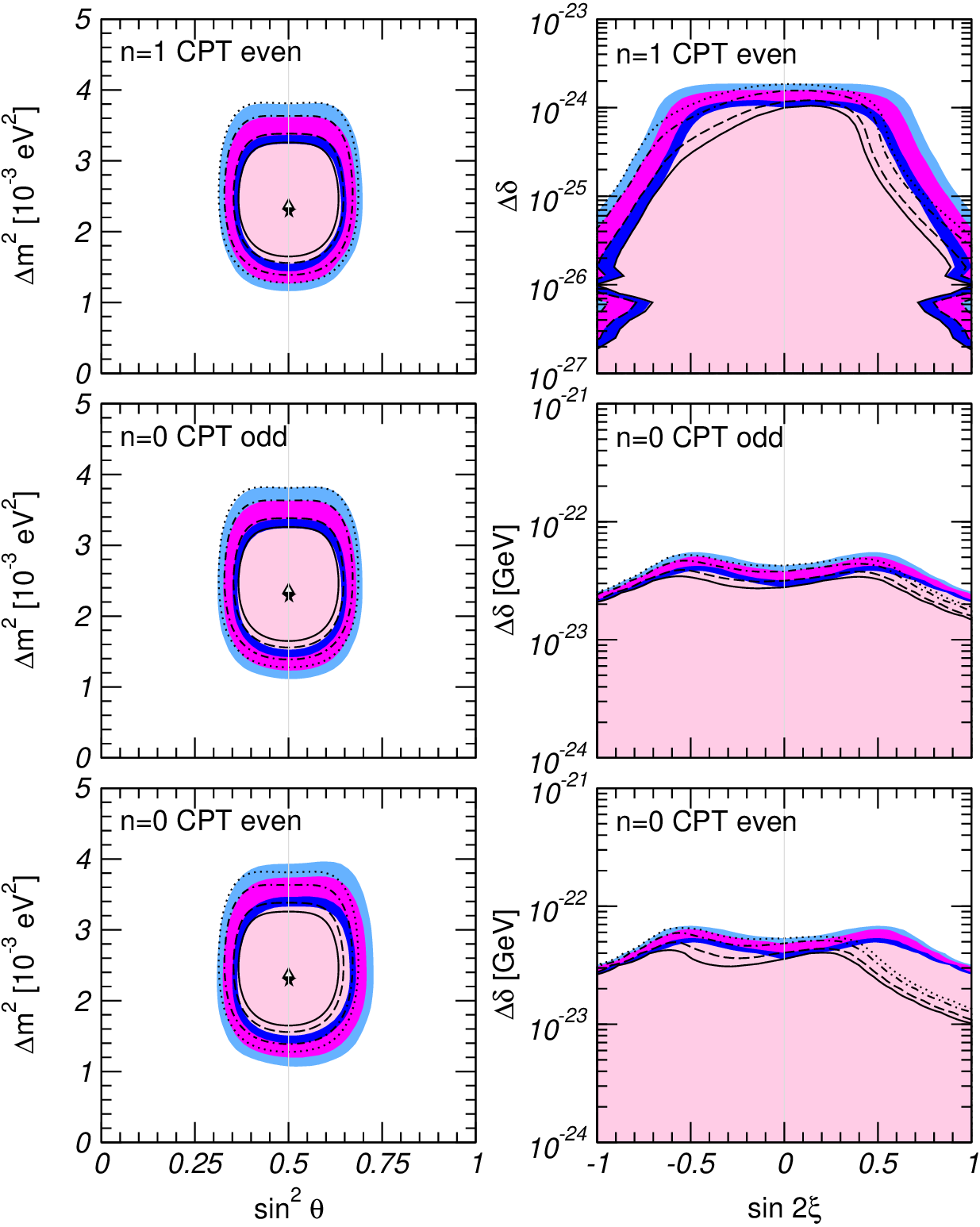}
    \caption{\label{fig:atmnp}%
      Allowed parameter regions for the analysis of atmospheric and
      K2K data in presence of $\nu_\mu \to \nu_\tau$ oscillations and
      different NP effects as labeled in the figure. Each panel shows
      a two-dimensional projection of the allowed five-dimensional
      region after marginalization with respect to the three
      undisplayed parameters. The different contours correspond to the
      two-dimensional allowed regions at 90\%, 95\%, 99\% and
      $3\sigma$ CL. The filled areas in the left panels show the
      projected two-dimensional allowed region on the oscillation
      parameters $\Dmq$--$\sin^2\theta$ plane. The best fit point is
      marked with a star. For the sake of comparison we also show the
      lines corresponding to the contours in the absence of new
      physics and mark with a triangle the position of the best fit
      point. The results are shown for the chosen relative sign
      $\sigma_n^+ = +1$; for $\sigma_n^+ = -1$ the corresponding
      region would be obtained by $\sin^2\theta \to 1-\sin^2\theta$.
      The regions on the right panels show the allowed values for the
      parameters characterizing the strength and mixing of the NP. The
      full regions corresponds to arbitrary values of the phase
      $\eta_n$ while the lines correspond to the case $\eta_n \in
      \{0,\,\pi\}$.}
\end{figure}

\begin{figure}
    \includegraphics[width=5in]{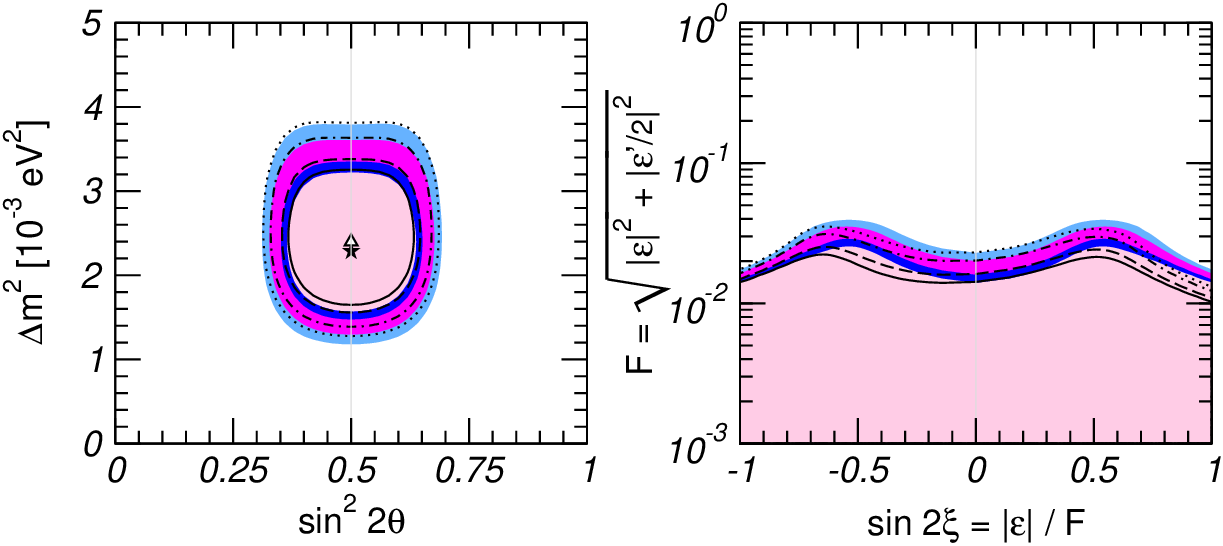}
    \caption{\label{fig:nsireg}%
      Same as Fig.~\ref{fig:atmnp} for the case of $\Delta m^2$-OSC+NSI.} 
\end{figure}

\begin{figure}
    \includegraphics[width=5in]{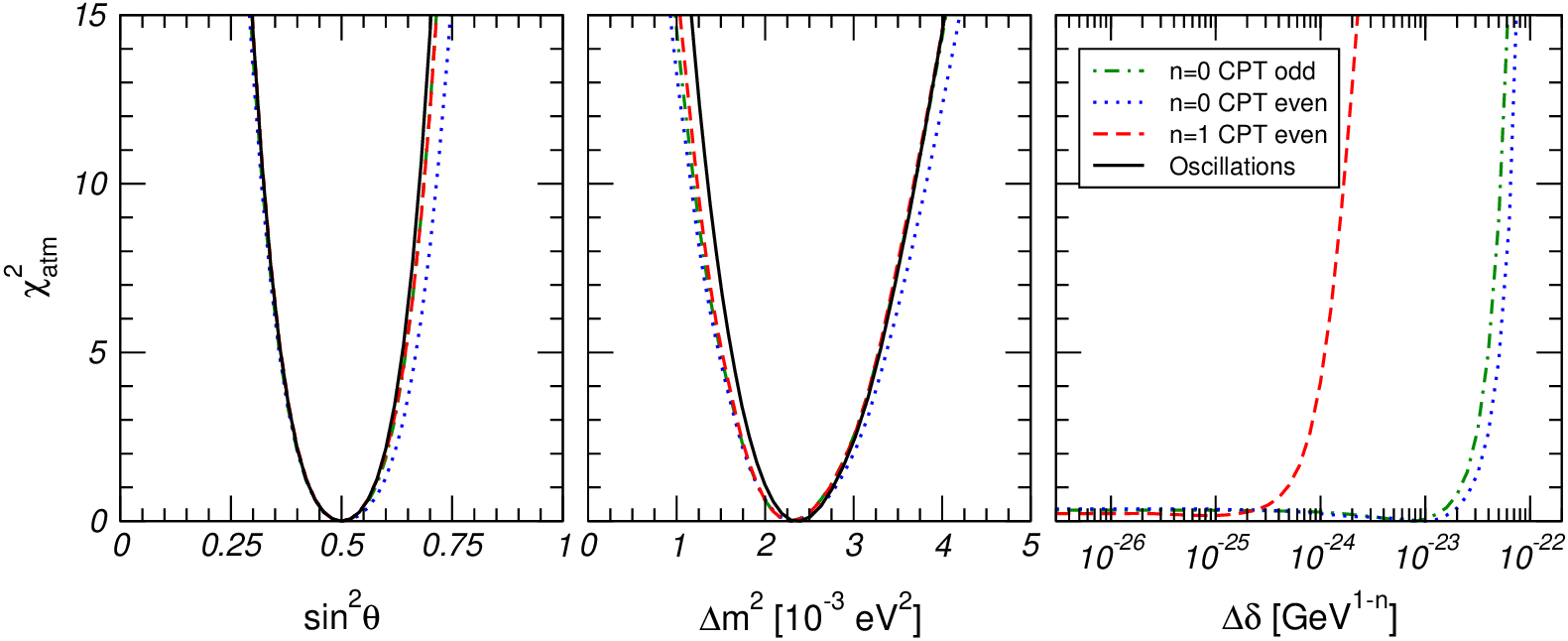}
    \caption{\label{fig:proj}%
      Dependence of $\Delta\chi^2$ on the oscillation parameters
      $\Dmq$, $\sin^2\theta$ and on the NP strength parameter $\Dlt_n$
      for different NP scenarios. The full line corresponds to pure
      $\nu_\mu \to \nu_\tau$ $ \Dmq$-OSC. The dashed, dotted and
      dot-dashed lines correspond to different $\Dmq$-OSC+NP scenarios
      as labeled in the figure. The figure is shown for
      $\sigma_n^+ = +1$. As described in the previous section the results
      hold for $\sigma_n^+ = -1$ with the exchange $\sin^2\theta \to 1 -
     \sin^2\theta$ (see discussion around Eq.~\eqref{eq:sign}).
      The individual $3\sigma$ bounds in
      Table~\ref{tab:ranges} can be read from the corresponding panel
      with the condition $\Delta\chi^2 \leq 9$.}
\end{figure}

\begin{figure}
    \includegraphics[width=4.in]{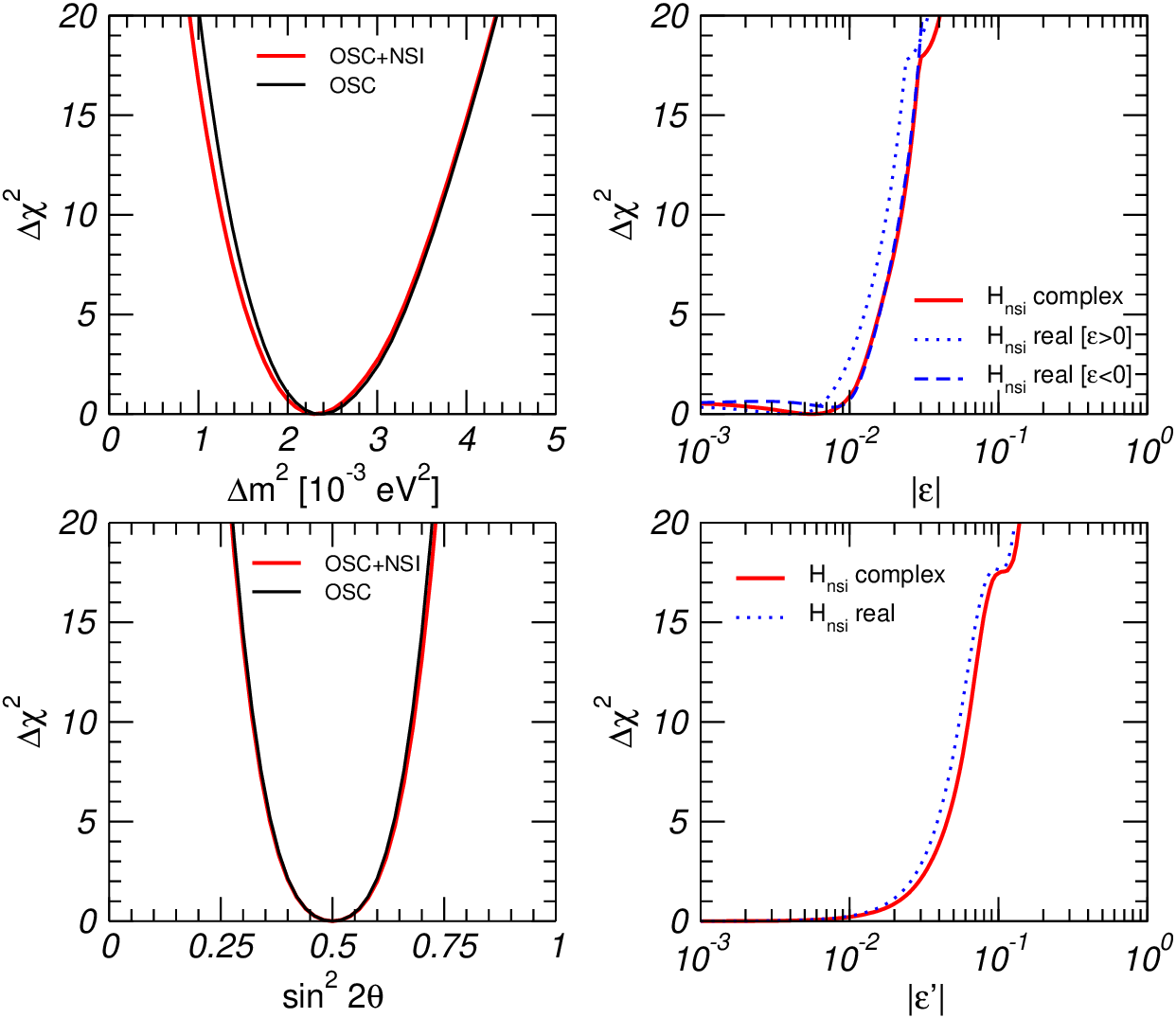}
    \caption{\label{fig:nsiproj}%
      Dependence of $\Delta\chi^2$ on the oscillation parameters
      $\Dmq$, $\sin^2\theta$ and on the NP strength parameters 
      for the case of $\Delta m^2$-OSC+NSI. }
\end{figure}

We now describe the results of our $\chi^2$ analysis of the standard
$\Dmq$-OSC+NP scenarios.  As discussed in the previous section the
analysis contains five parameters: $\Dmq$, $\theta$, $\Dlt_n$, $\xi_n$
and $\eta_n$ (or $\Epp$, $|\Eps|$ and $\arg(\Eps)$ for NSI).  Our
results are summarized in Figs.~\ref{fig:atmnp}, \ref{fig:nsireg},
\ref{fig:proj} and \ref{fig:nsiproj}, where we show different
projections of the allowed 5-dimensional parameter space.

In Figs.~\ref{fig:atmnp} and \ref{fig:nsireg} we plot two-dimensional
projections of the allowed parameter region for the analysis of
atmospheric and K2K data in presence of $\nu_\mu \to \nu_\tau$
oscillations and different NP effects parametrized in the form
Eq.~\eqref{eq:hamil}. The corresponding results for the case of NSI
are presented in Fig.~\ref{fig:nsireg}.  The regions in each panel are
obtained after marginalization of $\chi^2$ with respect to the three
undisplayed parameters and they are defined at 90\%, 95\%, 99\% and
$3\sigma$ CL for 2 d.o.f.\ ($\Delta\chi^2 = 4.61$, $5.99$, $9.21$ and
$11.83$, respectively).

The left panels in Figs.~\ref{fig:atmnp} and \ref{fig:nsireg} show the
projection of the allowed region on the oscillation parameters
$\Dmq$--$\sin^2\theta$ plane. The best fit point is marked with a
star. For the sake of comparison we also show the lines corresponding
to the contours of the allowed regions for pure $\Dmq$-OSC and mark
with a triangle the position of the best fit point. The results are
shown for the chosen relative sign $\sigma_n^+ = +1$. For $\sigma_n^+
= -1$ the corresponding region would be obtained by $\sin^2\theta \to
1 - \sin^2\theta$. 

The regions on the right panels of Fig.~\ref{fig:atmnp} and 
\ref{fig:nsireg} show the allowed values for the parameters
characterizing the strength and mixing of the NP. The full regions
correspond to arbitrary values of the relative phase $\eta_n$ (or
equivalently to complex $\Eps$ parameter for the NSI case) while the
lines show the results for real relative phase $\eta_n \in
\{0,\,\pi\}$ (which for NSI corresponds to $\Eps$ real and either
positive or negative, respectively).  For this second case we show the
allowed region for $-1 \leq \sin 2\xi_n\leq 1$ where for $\sigma_n^+ =
+1$ the sector with $-1 \leq \sin 2\xi_n\leq 0$ and $0\leq \sin
2\xi_n\leq 1$ correspond to $\eta_n = \pi$ and $\eta_n = 0$,
respectively, while the opposite holds for $\sigma_n^+ = -1$. As
discussed in the previous section, for the case of arbitrary phase
$\eta_n$ the full mixing parameter space can be covered by $0\leq
\xi_n \leq \pi/4$, which translates into the symmetry of the allowed
region around $\xi_n = 0$.

Several comments are in order. First, from the figures we see that the
best fit point for the global $\Dmq$-OSC+NP scenarios is always very
near the best fit point of pure $\Dmq$-OSC
\begin{equation} \label{bestfit}
    \Dmq_\text{best} = 2.2\times 10^{-3}~\eVq
    \qquad \sin^2\theta_\text{best} = 0.5 \,.
\end{equation}
In other words, the data does not show any evidence of presence of NP
even as a sub-dominant effect. Second, in agreement with SK
analysis~\cite{skatmlast}, we find that with the inclusion of the
three-dimensional atmospheric fluxes and improved cross sections as
well as with the reanalyzed data points from SK, the best fit point
and corresponding allowed regions from the atmospheric+K2K neutrino
analysis is shifted to slightly lower values of $\Dmq$ compared to our
previous analysis corresponding to the same data set~\cite{ourthree}.
Third, the figures illustrate the robustness of the allowed ranges of
mass and mixing derived from the analysis of atmospheric and K2K data
under the presence of these generic NP effects. Fourth the analysis
allow us to derive well-defined upper bounds on the NP strength. 

From Fig.~\ref{fig:atmnp} we see that the bounds on the NP strength
parameter $\Dlt_n$ tightens for larger values of $\xi_n$, being this
effect stronger for NP effects leading to sub-dominant oscillations
with stronger energy dependence. In particular, for $n=1$ the bound on
$\Dlt_n$ for $\sin^2 2\xi_n=1$ is a factor $\sim 50$ stronger than
that for $\xi_n=0$, while for $n=0$ the variation of the bound on
$\Dlt_n$ with $\xi_n$ is at most a factor $\sim 3$. This behaviour can
be qualitatively understood by studying the modification of the
oscillation probability at the best fit point of oscillations,
$\Dmq_\text{best} = 2.2\times 10^{-3}~\eVq$ and $\sin^2
2\theta_\text{best} = 1$, due to NP effects:
\begin{equation} \begin{split} \label{eq:pmtb}
    \Delta P \equiv \frac{P^\text{$\Dmq$+NP}_{\mu\mu} 
      - P^{\Dmq}_{\mu\mu}}{P^{\Dmq}_{\mu\mu}} = \tan^2\phi_b
    & - \frac{1 + 2 R_{n,b}\, \sin 2\xi_n \cos\eta_n 
      + R_{n,b}^2\, \sin^2 2\xi_n}
    {1 + 2 R_{n,b}\, \sin 2\xi_n \cos\eta_n + R_{n,b}^2} 
    \\[2mm]
    & \hspace{5mm} \times \frac{\sin^2 \left(\phi_b \sqrt{1 
	+ 2 R_{n,b}\, \sin 2\xi_n \cos\eta_n + R_{n,b}^2} \right)}
    {1 - \sin^2\phi_b}
\end{split} \end{equation}
where $\phi_b = 2.8~ (L / 10^3~\text{km})~ (\text{GeV} / E)$ is the
$\Dmq$ oscillation phase at the best fit point and $R_{n,b} = 0.91
\times 10^{21}~ (\Dlt_n / \text{GeV}^{1-n})~ (E / \text{GeV})^{n+1}$
is the ratio of NP to the standard oscillation contributions evaluated
at the best fit point of oscillations.

From Eq.~\eqref{eq:pmtb} we find that as long as $\phi_b$ is small the
dependence of $\Delta P$ on $R_{n,b}$ (and consequently on $\Dlt_n$)
is stronger for larger values of $|\sin 2\xi_n|$, which explains the
tightening of the bound on $\Dlt_n$. This behaviour was found in
Ref.~\cite{fogli1} for the case with $n=1$.

However, it is worth noticing that the characteristic value of
$\phi_b$ for which NP effects are relevant depends on $n$ since as $n$
increases the effect becomes important only for higher values of $E$
(see Fig.~\ref{fig:zenith}). As a consequence, the characteristic
$\phi_b$ for $n=0$ is larger than for $n=1$. Numerical inspection of
Eq.~\eqref{eq:pmtb} also shows that the variation of the dependence of
$\Delta P$ on $R_{n,b}$ with $\sin 2\xi_n$ decreases as $\phi_b$
increases. This explains the milder dependence of the bound on
$\Dlt_n$ with the mixing angle $\sin 2\xi_n$ for $n=0$ as compared
with the $n=1$ case.
The figure also illustrates that imposing that the Hamiltonian is real
does not substantially affect these conclusions.

The same arguments apply to the results for NSI in
Fig.~\ref{fig:nsireg}. In particular one sees that, as expected, the
results for NSI are very similar to to those derived for the $n=0$
CPT-odd scenario with the identification in Eq.~\eqref{eq:parnsi},
$\langle \Delta\rho_{\rm NSI}\rangle\sim 7\times 10^{-22} {\cal F}$
GeV.

More quantitative conclusions on the robustness of the derived ranges
for the oscillation parameters and on the bounds on the NP strength
can be obtained from Figs.~\ref{fig:proj} and~\ref{fig:nsiproj} where
we plot the marginalized $\Delta\chi^2$ as a function of the
oscillation parameters, $\Dmq$ and $\sin^2\theta$, and of the NP
strength parameters, for different NP scenarios as labeled in the
figures. 

In Table~\ref{tab:ranges} we list the $3\sigma$ allowed ranges for
$\Dmq$ and $\sin^2\theta$. We read that the derived ranges are robust
under the presence of these generic forms of NP whose only effect is
slightly enlarging the allowed range of $\Dmq$ by $\lesssim 15$\%, and
the lower bound on $\sin^22\theta$ by $\lesssim 7$\% at the $3\sigma$
level. We have verified that these conclusions hold for
other scenarios characterized by different values of $n$.

In terms of specific forms of NP the bounds on $\Dlt_n$ imply that
ATM+K2K limit the possible VLI in the $\nu_\mu$--$\nu_\tau$ sector via
CPT-even effects to
\begin{equation} \label{eq:VLIlim}
    |\delta v| \leq 8.1\times 10^{-25} ~ (1.6\times 10^{-24})
\end{equation}
and the possible VEP is constrained to
\begin{equation} \label{eq:VEPlim}
    |\phi\, \Delta \gamma| \leq 4.0\times 10^{-25} ~ (8.0\times 10^{-25})
\end{equation}
at 90\% ($3\sigma$), improving by a factor $8$ the
previous derived limits in these scenarios~\cite{fogli1}. We also find
that in the $\nu_\mu$--$\nu_\tau$ the VLI via CP-odd effects is
constrained to
\begin{equation} \label{eq:CPTlim}
    |\delta b| \leq 3.2\times 10^{-23} ~ (5.0\times 10^{-23})~\text{GeV}
\end{equation}
at 90\% ($3\sigma$), while non-universality of the neutrino couplings
to a torsion field
\begin{equation} \label{eq:torsionlim}
    |Q \,\delta k| \leq 4.0\times 10^{-23} ~ (6.3\times 10^{-23})~\text{GeV}
\end{equation}
at 90\% ($3\sigma$).

Finally for the case of non-standard neutrino interactions we find the
90\% ($3\sigma$) bounds
\begin{equation} \begin{split} \label{eq:nsilim}
    (-0.021)~{-0.013} \leq \Eps \leq 0.010~(0.017) \qquad
    & |\Epp|\leq 0.029~(0.052) \qquad
    \text{($\mathbf{H}_\text{NSI}$ real)},
    \\
    |\Eps|\leq 0.013~(0.021) \qquad
    & |\Epp|\leq 0.034~(0.060) \qquad
    \text{($\mathbf{H}_\text{NSI}$ complex)}
\end{split} \end{equation}
where the upper limits correspond to the case of real NSI and
the lower ones to the general case of complex $\Eps$.
These limits complement and update the previously derived bounds 
in Refs.~\cite{Fornengo:2001pm, lasteps}.

\section{Conclusions}
\label{sec:conclu}

In this work we have studied the robustness of our present
determination of the neutrino masses and mixing from the analysis of
the atmospheric and K2K data under the presence of some new physics
effects in the $\nu_\mu$--$\nu_\tau$ sector. In particular, we have
performed a global analysis to atmospheric and K2K data for scenarios
where vector-like or tensor-like new physics interactions affect the
neutrino evolution together with the standard $\Dmq$-mixing effect.

We have concluded that the data does not show any evidence of these
new physics effects even at the sub-dominant level. As a consequence
the derived range of oscillation parameters is robust under the
presence of those unknown effects. The quantification of this
statement is shown in Figs.~\ref{fig:proj} and~\ref{fig:nsiproj} and
in Table~\ref{tab:ranges}, from which we read that inclusion of these
new physics effects can at most enlarge the allowed range of $\Dmq$ by
$\lesssim 15$\% and relax the lower bound on $\sin^22\theta$ by
$\lesssim 7$\% at the $3\sigma$ level.

From the analysis we have also derived upper bounds on the strength of
the new physics effects in the $\nu_\mu$--$\nu_\tau$ sector. In
particular we show in Eqs.~\eqref{eq:VLIlim} and \eqref{eq:CPTlim} the
bound on the possible violation of Lorentz Invariance via CPT-even and
CPT-odd effects in the neutrino evolution respectively. The constraint
on the violation of the equivalence principle (VEP) due to non
universal coupling of the neutrinos to gravitational potential is
given in Eq.~\eqref{eq:VEPlim}, while bounds on non-universal
couplings of the neutrino to a torsion field are displayed in
Eq.~\eqref{eq:torsionlim}. The constraints on non-standard neutrino
interactions with matter are shown in Eq.~\eqref{eq:nsilim}.

\begin{table}
    \newcommand{\E}[2]{\ensuremath{{#1}\times 10^{#2}}}
    \begin{tabular}{|c|c|c|c|c|c|}
	\hline
	&  $\Dmq$-OSC & \multicolumn{4}{c|}{$\Dmq$-OSC+NP} \\
	\hline
	& & $n=1$ CPT-even & $n=0$ CPT-even & $n=0$ CPT-odd  &NSI  \\
	\hline
	$\Dmq$ [$10^{-3}~\eVq$] & 1.4--3.6   & 1.3--3.6         
      & 1.2--3.7         & 1.2--3.6       & 1.3--3.6        \\
	$\sin^2\theta$          & 0.33--0.67 & 0.33--0.68       & 0.33--0.71       & 0.33--0.68   &  0.33--0.67    \\
	$\Dlt_n$ [GeV$^{n+1}$]  & ---        & $< \E{1.6}{-24}$ & $< \E{6.3}{-23}$ & $< \E{5.0}{-23}$  & ${\cal F}\leq 0.035$\\
	\hline
    \end{tabular}
    \caption{\label{tab:ranges}%
      Individual $3\sigma$ ranges (with 1 d.o.f.) for the oscillation
      parameters $\Dmq$ and $\sin^2\theta$ for the different
      oscillation plus NP scenarios and $3\sigma$ bound on the NP
      strength parameters. The allowed range of
      $\sin^2\theta$ corresponds to $\sigma_n^+ = +1$. For $\sigma_n^+
      = -1$ the corresponding range would be obtained by $\sin^2\theta
      \to 1 - \sin^2\theta$.}
\end{table}

\acknowledgments

We are particularly indebted to M. Honda for providing us with their
new 3-dimensional atmospheric neutrino fluxes.  This work was
supported in part by the National Science Foundation grant PHY0098527.
MCG-G is also supported by Spanish Grants No FPA-2001-3031 and
CTIDIB/2002/24.

\appendix

\section{Statistical treatment of atmospheric data}

We summarize here our updated statistical analysis of the atmospheric
data. For convenience we have adopted the \emph{pull} method used
previously by the SK Collaboration~(see for instance
Refs.~\cite{Kameda,skthesis2} for details on their latest analysis)
and by the Bari group~\cite{Fogli:2002pt,fogli2}.  There are however
some technical differences which we describe next.

The basic idea of the pull method consists in parametrizing the
systematic errors and the theoretical uncertainties in terms of a set
of variables $\{\xi_i\}$, called \emph{pulls}, which are then treated
on the same footing as the other parameters of the model. The $\chi^2$
function can be decomposed into the sum of two parts:
\begin{equation} \label{eq:chisq-full}
    \chi^2(\vec\omega, \vec\xi) = 
    \chi^2_\text{data}(\vec\omega, \vec\xi) +
    \chi^2_\text{pulls}(\vec\xi),
\end{equation}
where $\vec\omega$ denotes the parameters of the model,
$\chi^2_\text{data}$ is the usual term describing the deviation of the
experimental results from their theoretical predictions, and the extra
term $\chi^2_\text{pulls}$ provides proper penalties to account for
deviations of the systematics and the theoretical inputs from their
nominal value. It is convenient to define the pulls in such a way 
that for each source
of systematics or theoretical input $i$ the value $\xi_i=0$
corresponds to the ``expected value'' reported by the collaboration or
predicted by the theory, and $\xi_i=\pm 1$ corresponds to a $1\sigma$
deviation. 

For the Super-Kamiokande experiment $\chi^2_\text{pulls}(\vec\xi)$ can
be properly written as a positive quadratic function of $\xi_i$.  
The interpretations of the pulls is particularly transparent 
if the sources of uncertainties are selected to be independent 
of each other. In this case the pulls are uncorrelated and 
the expression of $\chi^2_\text{pulls}$ is very simple:
\begin{equation}
    \chi^2_\text{pulls}(\vec\xi) =\sum_{i} \xi_i^2 \,.
\end{equation}

In its original formulation, the set of pulls selected by
Super-Kamiokande~\cite{Kameda} did not verify this condition and a
correlation matrix between the selected pulls (provided by SK
collaboration from their MC simulation) had to be included.  In our
analysis, however, we have identified the dominant independent sources
of systematic uncertainties in SK analysis, and we use them as the
basis for our pulls. We have characterized the theoretical and
systematic uncertainties in terms of 18 independent sources of error:
4 to parametrize the theoretical uncertainties associated to the
atmospheric fluxes (which we describe in Sec.~\ref{app:pullflux}), 6 for
the theoretical uncertainties in the interaction cross sections (given
in Sec.~\ref{app:pullcs}) and 8 sources of experimental systematic errors
(described in Sec.~\ref{app:pullsys}). To the point to which the
comparison is possible, this seems close to the approach followed by
Super-Kamiokande in their latest analysis~\cite{skthesis2}.  

The form of $\chi^2_\text{data}$ depends on the expected distribution
of the experimental results.  Under the standard approximation of
Gaussian distribution, we have:
\begin{equation}
    \chi^2_\text{data}(\vec\omega, \vec\xi) = \sum_n
    \left(
	\frac{R_n^\text{th}(\vec\omega, \vec\xi)
	  - R_n^\text{ex}}{\sigma_n^\text{stat}}
    \right)^2
\end{equation}
where $R_n^\text{th}$ ($R_n^\text{ex}$) is the ratio between the
expected (observed) number of events and the theoretical Monte Carlo
for the case of no oscillations.  Note that the dependence of
$\chi^2_\text{data}$ on both the parameters $\vec\omega$ and the pulls
$\vec\xi$ is entirely through $R_n^\text{th}(\vec\omega, \vec\xi)$. In
the pull approach, $\vec\omega$ and $\vec\xi$ play a very similar
role, and in principle should be treated in the same way.  However,
for the Super-Kamiokande experiment the bounds on $\vec\xi$ implied by
$\chi^2_\text{pulls}$ are in general significantly stronger than those
implied by $\chi^2_\text{data}$, and it is therefore a good
approximation to retain the dependence of $\chi^2_\text{data}$ on
$\vec\xi$ only to the lowest orders. This is done by expanding
$R_n^\text{th} (\vec\omega, \vec\xi)$ in powers of $\xi_i$ up to the
first order:
\begin{equation} \label{eq:approx}
  R_n^\text{th}(\vec\omega, \vec\xi) \approx
  R_n^\text{th}(\vec\omega)
  \left[ 1 + \sum_{i=1}^{18} \pi_n^i(\vec\omega)\, \xi_i \right],
  \quad \text{where} \quad 
  \left\{ \begin{aligned}
    R_n^\text{th}(\vec\omega) &\equiv R_n^\text{th}(\vec\omega, 0), \\[1mm]
    R_n^\text{th}(\vec\omega) \, \pi_n^i(\vec\omega) & \equiv
    \left. \frac{\partial R_n^\text{th}(\vec\omega,
	  \vec\xi)}{\partial \xi_i} \right|_{\vec\xi=0}.
    \end{aligned}\right.
\end{equation}
It is easy to prove~\cite{Fogli:2002pt} that under the
approximation~\eqref{eq:approx} Eq.~\eqref{eq:chisq-full} is
mathematically equivalent to the usual covariance definition of the
$\chi^2$ which we employed before in Refs.~\cite{ourthree,ourold}.
Thus the small difference in the results are not due to the different
statistical treatment, but to differences either in the input
parameters or in the updated values used for the systematic and
theoretical uncertainties.

Furthermore within the present precision one can safely neglect the
dependence of $\pi_n^i$ on the neutrino parameters $\vec\omega$. With
this approximation, we can write:
\begin{equation} \label{eq:chisq-work}
    \chi^2(\vec\omega) = \min_{\vec\xi} 
    \left[ {\displaystyle \sum_{n=1}^{55}} \left( 
	\frac{R_n^\text{th}(\vec\omega)
	\left[ 1 + \sum_i \pi_n^i\, \xi_i \right]
	- R_n^\text{ex}}
	{\sigma_n^\text{stat}} \right)^2 
	+ \sum_{i=1}^{18} \xi_i^2
    \right]
\end{equation}
where we have introduced the function $\chi^2(\vec\omega) =
\min_{\{\xi_i\}} \chi^2(\vec\omega,\vec\xi)$. It is clear from
Eq.~\eqref{eq:chisq-work} that in the present approach the systematic
and theoretical uncertainties are completely characterized by the set
of quantities $\{\pi_n^i\}$, which describe the strength of the
``coupling'' between the pull $\xi_i$ and the observable
$R^\text{th}_n$. 

In the rest of this section we will discuss in detail
how we have parametrized and taken into account the various sources of
uncertainty and list the derived values for $\{\pi_n^i\}$. 

\subsection{Flux uncertainties}
\label{app:pullflux}

\begin{table}
  \catcode`?=\active \def?{\hphantom{0}}
  \catcode`*=\active \def*{\times}
  \newcommand{\pul}[1]{\xi^\text{flux}_\text{#1}}
  \newcommand{\D}[2]{(#1,\> #2)}
  \newcommand{\zero}{\text{---}}
  \setlength{\extrarowheight}{-1pt}
  \begin{tabular}{|>{~}lr<{~}|CRCR|}
    \hline
    Sample & Bin 
    & \pul{norm} & \multicolumn{1}{C}{\pul{tilt}}
    & \pul{ratio} & \multicolumn{1}{C|}{\pul{zenith}}
    \\
    \hline
    ~            & ?1 & 20\% & \D{-1.44}{-1.11}*5\% & \D{-2.5\%}{+2.5\%} & -0.9*5\% \\
    ~            & ?2 & 20\% & \D{-1.43}{-1.11}*5\% & \D{-2.5\%}{+2.5\%} & -0.7*5\% \\
    ~            & ?3 & 20\% & \D{-1.42}{-1.11}*5\% & \D{-2.5\%}{+2.5\%} & -0.5*5\% \\
    ~            & ?4 & 20\% & \D{-1.42}{-1.10}*5\% & \D{-2.5\%}{+2.5\%} & -0.3*5\% \\
    sub-GeV      & ?5 & 20\% & \D{-1.42}{-1.10}*5\% & \D{-2.5\%}{+2.5\%} & -0.1*5\% \\
    $\D{e}{\mu}$ & ?6 & 20\% & \D{-1.42}{-1.10}*5\% & \D{-2.5\%}{+2.5\%} & +0.1*5\% \\
    ~            & ?7 & 20\% & \D{-1.42}{-1.10}*5\% & \D{-2.5\%}{+2.5\%} & +0.3*5\% \\
    ~            & ?8 & 20\% & \D{-1.43}{-1.10}*5\% & \D{-2.5\%}{+2.5\%} & +0.5*5\% \\
    ~            & ?9 & 20\% & \D{-1.44}{-1.10}*5\% & \D{-2.5\%}{+2.5\%} & +0.7*5\% \\
    ~            & 10 & 20\% & \D{-1.46}{-1.10}*5\% & \D{-2.5\%}{+2.5\%} & +0.9*5\% \\
    \hline
    ~            & ?1 & 20\% & \D{+0.35}{+0.91}*5\% & \D{-2.5\%}{+2.5\%} & -0.9*5\% \\
    ~            & ?2 & 20\% & \D{+0.38}{+0.92}*5\% & \D{-2.5\%}{+2.5\%} & -0.7*5\% \\
    ~            & ?3 & 20\% & \D{+0.42}{+0.94}*5\% & \D{-2.5\%}{+2.5\%} & -0.5*5\% \\
    ~            & ?4 & 20\% & \D{+0.49}{+0.98}*5\% & \D{-2.5\%}{+2.5\%} & -0.3*5\% \\
    multi-GeV    & ?5 & 20\% & \D{+0.56}{+1.04}*5\% & \D{-2.5\%}{+2.5\%} & -0.1*5\% \\
    $\D{e}{\mu}$ & ?6 & 20\% & \D{+0.56}{+1.04}*5\% & \D{-2.5\%}{+2.5\%} & +0.1*5\% \\
    ~            & ?7 & 20\% & \D{+0.49}{+0.98}*5\% & \D{-2.5\%}{+2.5\%} & +0.3*5\% \\
    ~            & ?8 & 20\% & \D{+0.43}{+0.95}*5\% & \D{-2.5\%}{+2.5\%} & +0.5*5\% \\
    ~            & ?9 & 20\% & \D{+0.39}{+0.93}*5\% & \D{-2.5\%}{+2.5\%} & +0.7*5\% \\
    ~            & 10 & 20\% & \D{+0.35}{+0.90}*5\% & \D{-2.5\%}{+2.5\%} & +0.9*5\% \\
    \hline
    ~            & ?1 & 20\% & +1.75*5\% & \zero & -0.9*5\% \\
    stopping     & ?2 & 20\% & +1.72*5\% & \zero & -0.7*5\% \\
    $\mu$ events & ?3 & 20\% & +1.73*5\% & \zero & -0.5*5\% \\
    ~            & ?4 & 20\% & +1.76*5\% & \zero & -0.3*5\% \\
    ~            & ?5 & 20\% & +1.84*5\% & \zero & -0.1*5\% \\
    \hline
    ~            & ?1 & 20\% & +4.64*5\% & \zero & -0.95*5\% \\
    ~            & ?2 & 20\% & +4.34*5\% & \zero & -0.85*5\% \\
    ~            & ?3 & 20\% & +4.48*5\% & \zero & -0.75*5\% \\
    ~            & ?4 & 20\% & +4.43*5\% & \zero & -0.65*5\% \\
    thrugoing    & ?5 & 20\% & +4.68*5\% & \zero & -0.55*5\% \\
    $\mu$ events & ?6 & 20\% & +4.62*5\% & \zero & -0.45*5\% \\
    ~            & ?7 & 20\% & +4.61*5\% & \zero & -0.35*5\% \\
    ~            & ?8 & 20\% & +4.96*5\% & \zero & -0.25*5\% \\
    ~            & ?9 & 20\% & +5.01*5\% & \zero & -0.15*5\% \\
    ~            & 10 & 20\% & +5.22*5\% & \zero & -0.05*5\% \\
    \hline
  \end{tabular}
  \caption{\label{tab:fluxes}%
    Coupling factors $\pi_n^i$ of the flux pulls $\pul{norm}$,
    $\pul{tilt}$, $\pul{ratio}$ and $\pul{zenith}$ with the various
    observables. When the notation $\D{v_1}{v_2}$ is used (second and
    third column for contained events), the first number refer to
    $e$-like events and the second to $\mu$-like events.}
\end{table}

Flux uncertainties are theoretical uncertainties arising from our
limited knowledge of the atmospheric neutrino fluxes. Following 
Refs.~\cite{Kameda,skthesis2} we have parametrized them in terms of
four pulls: $\xi^\text{flux}_\text{norm}$,
$\xi^\text{flux}_\text{tilt}$, $\xi^\text{flux}_\text{ratio}$ and
$\xi^\text{flux}_\text{zenith}$. The corresponding coefficients
$\pi_n^i$ are listed in Table~\ref{tab:fluxes}.
\begin{itemize}
  \item $\xi^\text{flux}_\text{norm}$ is the pull associated to the 
    total normalization error, which we set to 
    $\sigma^\text{flux}_\text{norm}$=20\%~\cite{gaisser}.  

  \item $\xi^\text{flux}_\text{tilt}$ is a``tilt'' factor which
    parametrizes possible deviations of the energy dependence of the 
   atmospheric fluxes from the simple power law.
    Following Refs.~\cite{Kameda,skthesis2}, we define:
    \begin{equation}
	\Phi_\delta(E) = \Phi_0(E) \left( \frac{E}{E_0} \right)^\delta
	\approx \Phi_0(E) \left[ 1 + \delta \ln \frac{E}{E_0} \right]
    \end{equation}
    and assume an uncertainty on the factor $\delta$,
    $\sigma_\delta=5\%$~\cite{Kameda,skthesis2}. Also in analogy with 
    Refs.~\cite{Kameda,skthesis2} we have chosen $E_0 = 2$~GeV. We
    then calculate numerically the coefficients $\pi_n^\text{tilt}$ as
    follows: we compute the expected number of events for a given bin
    $N_n$ using $\Phi_\delta(E)$ for the central value of $\delta$ and
    for $\delta\pm\sigma_\delta$ and obtain the corresponding coupling
    $\pi^\text{tilt}_n$ as the relative change in $N_n$. The results
    reported in the second column of Table~\ref{tab:fluxes} are
    obtained neglecting the effect of oscillations. However we have
    verified that when the dependence of the $\pi_n^\text{tilt}$ on
    the neutrino oscillation parameters is properly taken into account
    we find very similar results. 
    
  \item $\xi^\text{flux}_\text{ratio}$ parametrizes the uncertainty on
    the $\nu_\mu / \nu_e$ ratio, which is assumed to be 
    $\sigma_{\mu/e}= 5\%$~\cite{gaisser, Kameda,skthesis2} and
    following Ref.~\cite{Kameda} we assign a coupling
    $\pi^{\mu/e}_{\mu} = -\pi^{\mu/e}_{e} = \sigma_{\mu/e}/2$.
    
  \item $\xi^\text{flux}_\text{zenith}$ describes the uncertainty
    on the zenith angle dependence, which we assume energy
    independent. As in Ref.~\cite{Kameda} we parametrize the coupling
    for this pull for the bin $n$ of a given sample as
    $\pi^\text{zenith}_n=5\% \; \langle \cos\theta \rangle_n$. This
    means that this uncertainty can induce an error in the up/down
    asymmetry of events which we conservatively take to be 5\%. In
    Ref.~\cite{Kameda} the assumed up/down uncertainty was smaller
    (2.5\%) and a separate zenith-pull was introduced for the
    horizontal-to-vertical ratio uncertainty of 2\%. We have verified
    that within the present precision both parametrizations of the
    uncertainties in the zenith angle distribution give very similar
    results. 
\end{itemize}

\subsection{Cross-section uncertainties}
\label{app:pullcs}

\begin{table}
  \catcode`?=\active \def?{\hphantom{0}}
  \newcommand{\zero}{\text{---}}
  \newcommand{\pul}[2]{\xi^\text{#1}_\text{#2}}
  \begin{tabular}{|>{~}l<{~}|CC|CC|CC|}
    \hline
     Sample & \pul{QE}{norm}  & \pul{QE}{ratio}
    & \pul{$1\pi$}{norm} & \pul{$1\pi$}{ratio}
    & \pul{DIS}{norm}    & \pul{DIS}{ratio} \\
    \hline
    sub-GeV $e$      & 11.3\% & -0.19\% & ?3.2\% & -0.10\% & ?0.5\% & -0.01\% \\
    sub-GeV $\mu$    & 11.3\% & +0.19\% & ?3.2\% & +0.11\% & ?0.5\% & +0.01\% \\
    multi-GeV $e$    & ?6.1\% & -0.20\% & ?5.0\% & -0.13\% & ?3.9\% & -0.49\% \\
    multi-GeV $\mu$  & ?2.1\% & +0.07\% & ?5.2\% & +0.14\% & ?7.7\% & +0.98\% \\
    stopping $\mu$   & ?2.3\% & \zero   & ?1.4\% & \zero   & ?7.5\% & \zero   \\
    thrugoing $\mu$  & ?0.5\% & \zero   & ?0.2\% & \zero   & ?9.6\% & \zero   \\
    \hline
  \end{tabular}
  \caption{\label{tab:crsect}%
    Coupling factors $\pi^i_n$ of the cross-section pulls
    $\pul{QE}{norm}$, $\pul{QE}{ratio}$, $\pul{$1\pi$}{norm}$,
    $\pul{$1\pi$}{ratio}$, $\pul{DIS}{norm}$ and $\pul{DIS}{ratio}$
    with the various observables. The couplings are the same for all
    the bins in a given data sample.}
\end{table}

Cross section uncertainties are theoretical uncertainties associated
to our ignorance on the neutrino interaction cross section.  In our
calculation we follow the standard approach and consider separately
the contributions to the interaction cross section from the exclusive
channels of lower multiplicity: quasi-elastic scattering (QE), and
single pion production ($1\pi$), and include all additional channels
as part of the deep inelastic (DIS) cross section (also refer to as
multi-pion).  We neglect for simplicity coherent scattering on oxygen
and neutral-current interactions, which contribute only marginally to
the considered data samples. 

We assume that each of these three contributions to the cross sections
are subject to different sources of uncertainties which allow to 
consider the corresponding pulls as independent.  For each type of
neutrino interactions we introduce two pulls:
\begin{itemize}
  \item $\xi^\text{QE}_\text{norm}$, $\xi^{1\pi}_\text{norm}$,
    $\xi^\text{DIS}_\text{norm}$ describe the total normalization
    errors. We conservatively assume 
    $\sigma^{\sigma_\text{QE}}_\text{norm}=15$\% and
    $\sigma^{\sigma_{1\pi}}_\text{norm}=15$\%. For the normalization
    error of the DIS cross section we estimate
    $\sigma^{\sigma_\text{DIS}}_\text{norm}=15$\% for contained events
    and $\sigma^{\sigma_\text{DIS}}_\text{norm}=10$\% for upward-going
    muons from the spread of theoretical predictions arising from the
    use of different sets of nucleon structure functions. The relevant
    coefficients $\pi_n^i$ are listed in Table~\ref{tab:crsect}. They
    are obtained computing the relative change in the number of
    expected events in a given data sample arising from the use of
    either $\sigma_{i}$ or $\sigma_{i} \pm
    \sigma^{\sigma_\text{i}}_\text{norm}$ for each of the three
    contributions to the cross section.
    
  \item $\xi^\text{QE}_\text{ratio}$, $\xi^{1\pi}_\text{ratio}$,
    $\xi^\text{DIS}_\text{ratio}$ parametrize the uncertainty of the
    $\sigma_{i,\nu_\mu} / \sigma_{i,\nu_e}$ ratios. This error is
    relevant only for contained events, and it is much smaller than
    the total normalization uncertainty. The numbers listed in
    Table~\ref{tab:crsect} are obtained from Ref.~\cite{Kameda}.
\end{itemize}

\subsection{Systematic uncertainties}
\label{app:pullsys}

\begin{table}
  \catcode`?=\active \def?{\hphantom{0}}
  \newcommand{\pul}[1]{\xi^\text{sys}_\text{#1}}
  \newcommand{\zero}{\text{---}}
  \begin{tabular}{|>{~}l<{~}|CCCCCCCC|}
    \hline
    \multicolumn{1}{|c|}{Sample}
    & \pul{hadron} & \pul{$\mu/e$} & \pul{ring} & \pul{f-vol} & \pul{E-cal}
    & \pul{FC/PC} & \pul{track} & \pul{up-eff} \\
    \hline
    sub-GeV $e$     & -0.25\% & -1.1\% & -0.75\% & -0.3\% & -0.4\% & \zero  & \zero & \zero \\
    sub-GeV $\mu$   & +0.25\% & +1.1\% & +0.75\% & +0.3\% & +0.4\% & \zero  & \zero & \zero \\
    multi-GeV $e$   & -0.50\% & -1.6\% & -2.75\% & -0.5\% & -0.4\% & \zero  & \zero & \zero \\
    multi-GeV $\mu$ & +1.10\% & +1.6\% & +5.40\% & +1.4\% & +2.0\% & 2.85\% & \zero & \zero \\
    stopping $\mu$  & \zero   & \zero  & +0.30\% & +0.7\% & +0.3\% & \zero  & 6.4\% & 1.4\% \\
    thrugoing $\mu$ & \zero   & \zero  & +0.30\% & +0.7\% & +0.3\% & \zero  & 1.4\% & 1.0\% \\
    \hline
  \end{tabular}
  \caption{\label{tab:syst}%
    Coupling factors of the systematics pulls $\pul{hadron}$,
    $\pul{$\mu/e$}$, $\pul{ring}$, $\pul{f-vol}$, $\pul{E-cal}$,
    $\pul{PC-nrm}$, $\pul{FC/PC}$, $\pul{track}$ and $\pul{up-eff}$
    with the various observables. The coefficients are the same for
    all the bins in a given data sample.}
\end{table}

The systematics uncertainties of the Super-Kamiokande experiment are
derived from Tables 9.2, 9.3, 9.4 and 9.5 of Ref.~\cite{Kameda}. We
include in our calculations the following sources of systematics:
\begin{itemize}
  \item $\xi^\text{sys}_\text{hadron}$ is the pull for the uncertainty
    associated with the simulation of hadronic interactions;
    
  \item $\xi^\text{sys}_{\mu/e}$ is the pull for  the errors in the
    particle identification procedure;
    
  \item $\xi^\text{sys}_\text{ring}$ is the pull for the uncertainty
    coming from the ring-counting procedure;

  \item $\xi^\text{sys}_\text{f-vol}$ is the pull for the uncertainty
    in the fiducial volume determination;
    
  \item $\xi^\text{sys}_\text{E-cal}$ is the pull for the uncertainty
    in the energy calibration;
    
  \item $\xi^\text{sys}_\text{PC-nrm}$ is the pull for the relative
    normalization between partially-contained and fully-contained
    events.
    
  \item $\xi^\text{sys}_\text{track}$ is the pull for the uncertainty
    in the track reconstruction of upgoing muons;
    
  \item $\xi^\text{sys}_\text{up-eff}$ is the pull for the uncertainty
    in the detection efficiency of upgoing muons and the
    stopping-thrugoing separation.
\end{itemize}

\end{document}